\documentclass{amsart}
\usepackage{amsmath}
\usepackage{graphicx} 
\usepackage{amsfonts} 
\usepackage{amssymb}
\usepackage{float}
\usepackage{color}
\usepackage{mathrsfs}
\usepackage{epsfig}
\usepackage{subfig}
\usepackage{url}
\theoremstyle{plain}
\newtheorem{theorem}{Theorem}[section]

\newtheorem{lemma}[theorem]{Lemma}

\theoremstyle{definition}
\theoremstyle{remark}
\numberwithin{equation}{section}

\newcommand{\R}{\mathbb{R}}
\newcommand{\N}{\mathbb{N}}
\newcommand{\Z}{\mathbb{Z}}

\newcommand{\E}{\mathcal{E}}

\newcommand{\XLN}{\mathcal X_{\mathcal L}^N}
\newcommand{\XLNN}{\mathcal X_{\mathcal L}^{N'}}

\definecolor{viole}{RGB}{100,0,120}
\definecolor{darkor}{RGB}{220,110,0}

\usepackage{latexsym}
 
\title [Uniqueness of the minimizers for the Heitmann-Radin energy]{Classification of particle numbers\\
 with unique Heitmann-Radin minimizer }

\author[L. De Luca]
{L. De Luca}
\address[Lucia De Luca]{ Zentrum Mathematik - M7, Technische Universit\"at  M\"unchen, Boltzmannstrasse 3, 85748 Garching, Germany}
\email[L. De Luca]{deluca@ma.tum.de}

\author[G. Friesecke]
{G. Friesecke}
\address[Gero Friesecke]{ Zentrum Mathematik - M7, Technische Universit\"at  M\"unchen, Boltzmannstrasse 3, 85748 Garching, Germany}
\email[G. Friesecke]{gf@ma.tum.de}

\begin{document}
\large

\begin{abstract}
We show that minimizers of the Heitmann-Radin energy \cite{HR} are unique if and only if the particle number $N$ belongs to an infinite sequence whose first thirty-five elements are 1, 2, 3, 4, 5, 7, 8, 10, 12, 14, 16, 19, 21, 24, 27, 30, 33, 37, 40, 44, 48, 52, 56, 61, 65, 70, 75, 80, 85, 91, 96, 102, 108, 114, 120 (see the paper for a closed-form description of this sequence). The proof relies on the discrete differential geometry techniques introduced in \cite{DLF}.
\end{abstract}

\maketitle


\section{Introduction} 
A fundamental problem in statistical and solid mechanics is to explain theoretically why atoms at low temperature self-assemble into subsets of periodic lattices, and why these subsets exhibit specific polyhedral shapes. 

In previous studies, the specific shapes have been beautifully explained under two simplifications. First, one {\it assumes} crystallization, i.e. one restricts the admissible atomic positions to lattice sites. Second, one passes to a coarse-grained description in which atomistic energy minimization is replaced by minimization of an effective surface energy of the region $\Omega$ occupied by the atoms. Such a surface energy governing the shape $\Omega$ was first written down by Gibbs, and in modern notation has the form
\begin{equation} \label{gibbs}
   \int_{\partial\Omega} \varphi(\nu(x)) \, dA(x).
\end{equation}
Here $\Omega$ is an -- up to regularity requirements arbitrary -- subset of $\R^d$, $\nu(x)$ is the outward unit normal to $\Omega$ at the point $x$, $\varphi$ is a surface energy density which captures the fact that interfaces with certain orientations with respect to the crystal lattice are favoured over others, and $dA(x)$ is the usual area element (Hausdorff measure ${\mathcal H}^{d-1}$) on $\partial\Omega$. 
A fundamental result going back to Taylor \cite{Taylor} and Fonseca and M\"uller \cite{FM} states that minimizers of \eqref{gibbs} among sets of finite perimeter and fixed volume are unique up to translation, and given by a dilation of the {\it Wulff shape} 
$\{ x\in\R^d \, : \, x \cdot \nu(x) \le \varphi(x) \mbox{ for all  }x\in\partial\Omega\}$.
For a related uniqueness result at finite temperature in the context of the 2D Ising model see \cite{DKS}. 

Microscopically, the situation regarding uniqueness is much more subtle. Our goal in this note is to settle the uniqueness question completely in case of the two-dimensional Heitmann-Radin model, the perhaps simplest model describing the self-assembly of atoms into crystalline order and special shapes despite allowing arbitrary particle positions in $\R^2$. 

The Heitmann-Radin energy for a system of $N$ identical particles with positions $x_1,..,x_N\in\R^2$, introduced in \cite{HR}, is
\begin{equation}\label{energy}
\E_{HR}(X):=\frac 1 2 \sum_{1\le i< j\le N} V_{HR}(|x_i-x_j|),
\end{equation}
where here and below we abbreviate the position vector of all particles by $X\in\R^{2N}$ and the interaction potential $V_{HR}$ is given by
\begin{equation} \label{VHR}
V_{HR}(r)=\left\{\begin{array}{ll}
+\infty&\textrm{if } r<1\\
-1 &\textrm{if } r=1\\
0&\textrm{if }r>1.
\end{array}
\right.
\end{equation}
The potential \eqref{VHR} arises naturally from the Lennard-Jones potential $V(r)=r^{-2p}-2r^{-p}$ by passing to the limit $p\to\infty$ \cite{DLF}. In \cite{HR}, Heitmann and Radin proved the fundamental result that for any fixed $N\in\N$, the configurations $X$ minimizing $\E_{HR}$ are, up to rotation and translation, subsets of the triangular lattice
\begin{equation*}
\mathcal L=\{i\mathbf{e}+j\mathbf{f}\,:\,i,j\in\Z\},\quad\mathbf{e}=\left(\begin{array}{l} 1\\ 0\end{array}\right),\quad
\mathbf{f}=\left(\begin{array}{l} 1/2\\ {\sqrt 3}/2\end{array}\right).
\end{equation*}
Moreover the authors give an explicit formula for the ground state energy from which it is elementary to deduce examples of non-uniqueness. Such examples by no means contradicts the macroscopic uniqueness result for \eqref{gibbs}; indeed in \cite{AFS} the associated macroscopic surface energy density $\varphi$ is derived for the model \eqref{energy}, and it is shown that as the number of $N$ of particles gets large, the empirical measure $\mu_N$ associated with any sequence of microscopic minimizers converges after re-scaling to the characteristic function of the unique Wulff shape. 

\begin{figure}[H]
\begin{center}
\includegraphics[width=1.0\textwidth]{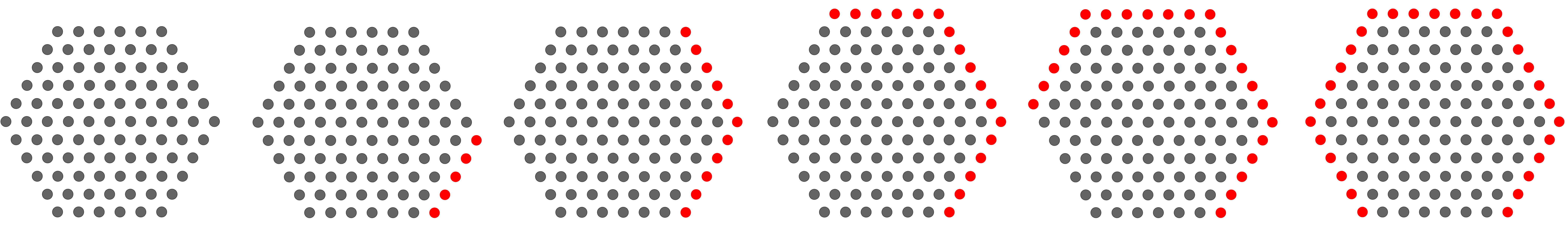}
\end{center}
\caption{
The uniqueness cases for $N=3s ^2+3 s+1+\ell$ with $0\le \ell\le 6 s$ and $s=5$. Starting from the left: $N=91$, $N=91+5=96$, $N=91+(5+1)\cdot 1+5=102$, $N=91+(5+1)\cdot 2+5=108$, $N=91+(5+1)\cdot 3+5=114$, $N=91+(5+1)\cdot 4+5=120$. The particles added to the regular hexagon on the left in order to obtain the unique minimizers are shown in red.}
\label{F:uniqueness}
\end{figure}

Further insight into the uniqueness question was recently achieved by Schmidt \cite{S}. He showed that microscopic regular hexagons (Figure \ref{F:uniqueness}, left picture) are the unique microscopic minimizer for their particle number, by re-writing the discrete energy in the form \eqref{gibbs} via introducing suitable tiles around each particle and applying the macroscopic uniqueness theorem for \eqref{gibbs}. Moreover in \cite{S} it is shown that for particle numbers $N$ exceeding those of a microscopic regular hexagon by $1$, the amount of nonuniqueness can be surprisingly large (for a refinement of this result see \cite{DPS}).

Here we completely settle the question of when microscopic uniqueness occurs. The result is as follows.
\begin{theorem} \label{uniqueness} 
Let $N\in\N$. The minimizers of $\E_{HR}$ among $N$-particle configurations are unique up to translation and rotation if and only if either
$$
 (a) \;\;\;  N=3s^2+3s+1
$$
for some $s\in\N\cup\{0\}$, or 
$$
  (b) \;\;\; N=3s^2+3s+1+(s+1)k+s
$$
for some $s\in\N\cup\{0\}$ and some $k\in\{0,1,2,3,4\}$.
\end{theorem}
An explicit list of the numbers between 1 and 120 satisfying (a) or (b) was given in the Abstract. Case (a) corresponds to regular hexagons (see the leftmost picture in Figure \ref{F:uniqueness}), recovering the uniqueness result of \cite{S}. The other uniqueness cases were neither previously conjectured, nor can they be established with the same method, as the microscopic minimizers are no longer regular hexagons and hence the associated continuum sets introduced in \cite{S} do not minimize the correspoding continuum energy \eqref{gibbs}. 

\begin{figure}[H]
\begin{center}
\includegraphics[width=0.5\textwidth]{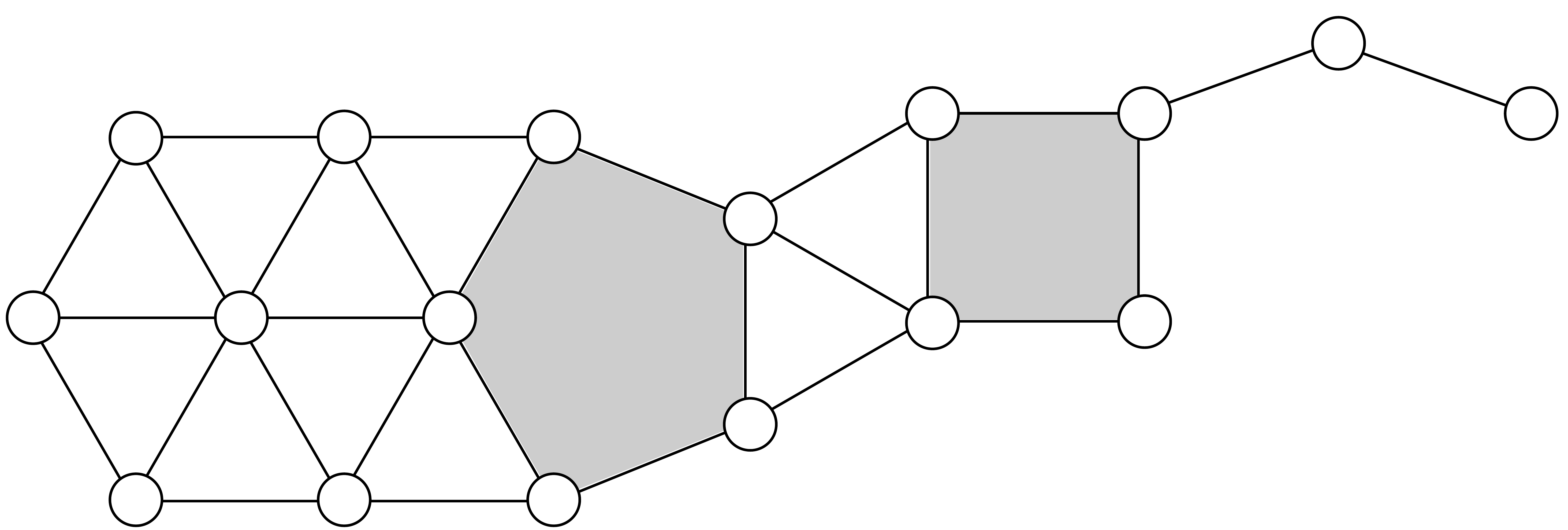}
\end{center}
\caption{Bond graph, defect measure and combinatorial perimeter of a particle configuration.\\
Particles are said to be connected by an edge (or bond) if their distance is $1$. For the bond graph above, the Euler characteristic $\chi$ is $1$ (since the graph is connected), the defect measure $\mu$ (see \eqref{defmu}) equals $3$ (since the graph contains a square and a pentagon), and the combinatorial perimeter equals $17$. The latter is because there are $13$ regular boundary edges, i.e. edges lying on the boundary of precisely one face, and $2$ wire edges, i.e. edges not lying on the boundary of any face, which must be counted twice. It follows that the right hand side of \eqref{energydecomp} equals $-3\cdot 17 + 17 + 3 + 3\cdot 1 = -28$, which indeed agrees with the Heitmann-Radin energy of the configuration.}\label{bondgraph}
\end{figure}

Our proof of Theorem \ref{uniqueness} relies on the discrete differential geometry approach introduced recently by us in \cite{DLF}. In \cite{DLF} we used this approach to give a new proof of the Heitmann-Radin crystallization theorem; here we employ it to settle the uniqueness question. This approach starts by associating, to each particle configuration $X=(x_1,..,x_N)$, its bond graph, a planar graph (see Figure \ref{bondgraph}) whose definition we recall here: vertices correspond to the particle positions $x_j$, edges to line segments $[x_j,x_k]$ connecting two particle positions of distance $1$, and faces to open bounded subsets of $\R^2$ which are nonempty, do not contain any point in $X$, and whose boundary is given by a cycle $\cup_{i=1}^k[x_{i-1},x_i]$ of edges for some points $x_0,..,x_k$ with $x_k=x_0$. 
A key result of \cite{DLF} needed in the proof of Theorem \ref{uniqueness} is the following geometric decomposition of the Heitmann-Radin energy on $N$-particle configurations: 
\begin{equation}\label{energydecomp}
\E_{HR}(X)=-3N+P(X)+\mu(X)+3\chi(X),
\end{equation}
where $\chi(X)$ is the Euler characteristic of the bond graph of $X$, $P(X)$ is its combinatorial perimeter as introduced in \cite{DLF}, namely the number of boundary edges with ``wire edges'' counted twice (see Figure \ref{bondgraph}), and
$\mu(X)$ is the defect measure
\begin{equation}\label{defmu}
\mu(X)=\sharp\textrm{quadrilaterals}+2\,\sharp\textrm{pentagons}+3\,\sharp\textrm{hexagons}+\ldots, 
\end{equation}
which can be viewed as a distance measure between the bond graph and vacancy-free subsets of the triangular lattice $\mathcal{L}$. See Figure \ref{bondgraph}. 

\section{Proof of the theorem}

\begin{figure}[H]
\begin{center}
\includegraphics[width=0.65\textwidth]{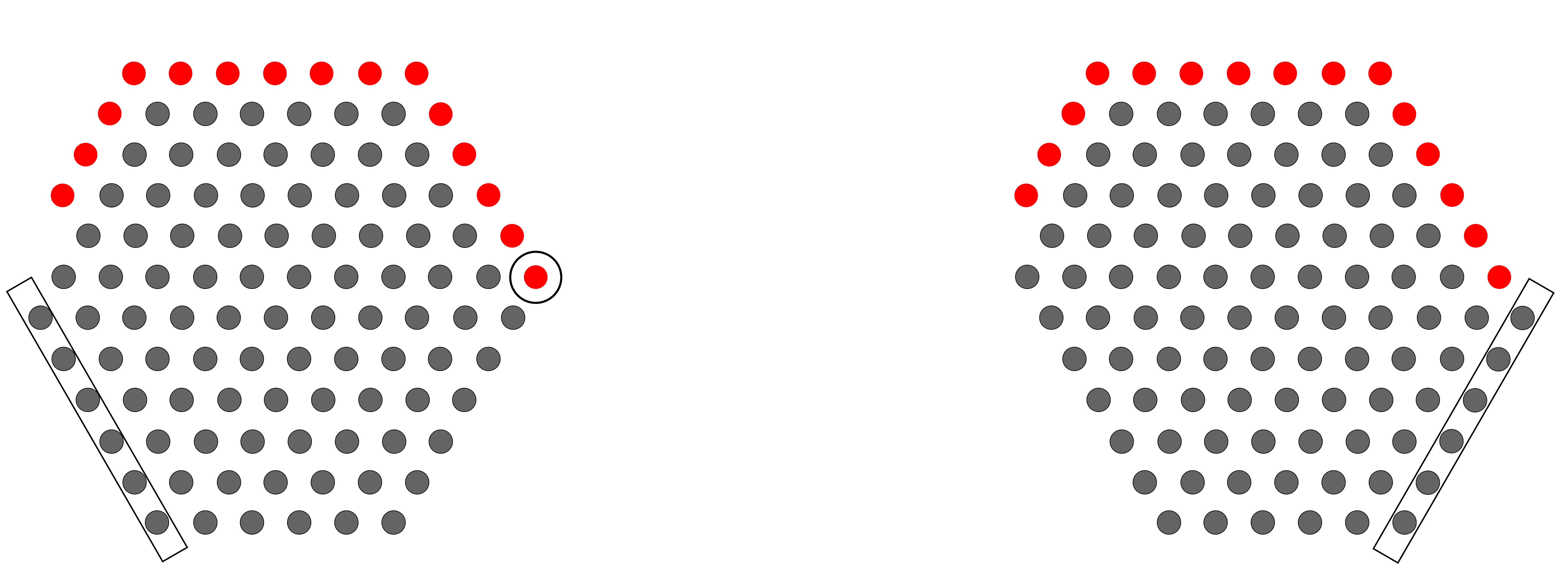}
\end{center}
\caption{Two minimizers of $\mathcal{E}_{HR}$ for $N=106=3\,s^2+3\,s+1+(s+1)k+j$, with $s=5$, $k=2$, and $j=3$.\\
Left: The ``canonical'' minimizer from \cite{HR, DLF}: starting from the grey particles constituting a regular hexagon with sidelength $s$, the remaining (red) particles are added by following a counterclockwise path around the hexagon starting from the encircled point.\\
Right: Another minimizer, obtained from the one on the left by moving the boxed segment of atoms. 
}
\label{F:counterex}
\end{figure}

We begin by showing non-uniqueness in the case when the particle number $N$ is not of the form (a) or (b) in Theorem \ref{uniqueness}. First notice that for such $N$, the ``canonical'' minimizer from \cite{HR, DLF} (see Figure \ref{F:counterex}, left picture) has a non-convex angle at the boundary. 

As in the Figure, let $k\in\{0,1,2,3,4,5\}$ be the number of complete sides of the canonical minimizer that were added to the central regular hexagon. When $k\le 2$, we can move the first not yet covered side of the central hexagon (boxed particles, left picture in Figure \ref{F:counterex}) on top of the last not yet covered side of the central hexagon (see Figure \ref{F:counterex}, right picture). This preserves $\mu=0$ and $\chi=1$, and does not change the perimeter, and hence, by \eqref{energydecomp}, yields another minimizer. When $k\ge 3$, we can instead move the first new side of the canonical minimizer obtained by covering the central hexagon on top of the third covered side, again leaving defect measure, Euler characteristic and perimeter -- and hence the Heitmann-Radin energy -- unchanged. Either way, the new minimizer is not a rotated translate of the canonical one. 

\vskip 2mm

Before moving to the proof of uniqueness when $N$ is of form (a) or (b), we introduce some notation (which was also used in \cite{DLF}).  
For an $N$-point configuration, i.e. $X\subset\R^2$ with $\sharp X = N$, endowed with the planar graph structure described in the Introduction, we say that $x\in X$ is a {\it boundary particle} if it is adjacent to a boundary edge, i.e. an edge lying on the boundary of at most one face. We denote the set of boundary particles by $\partial X$, and say that $X$ has simply closed polygonal boundary if the union of boundary edges forms a simply closed curve. 

Let us also recall the Heitmann-Radin crystallization theorem proved in \cite{HR} (see also \cite{DLF}). This result says that minimizers of the Heitmann-Radin energy \eqref{energy} among arbitrary $N$-particle configurations, i.e. $X\subset\R^2$ with $\sharp X = N$, belong -- up to translation and rotation and for $N\ge 3$ -- to the set of crystallized configurations
\begin{multline*}
\XLN:=\{X\subset\mathcal L \, : \, \sharp X = N, \mbox{ all faces of }X\mbox{ are triangles}, \\
X\textrm{ has simply closed polygonal boundary}\}.
\end{multline*}
This reduces the uniqueness question to uniqueness among configurations in $\XLN$, and together with \eqref{energydecomp} shows that
\begin{equation*}
    \min_{X\subset\R^2, \, \sharp X=N} \E_{HR}(X) = \min_{X\in\XLN}\E_{HR}(X)=-3N + 3 + \min_{X\in\XLN}P(X).
\end{equation*}
Moreover, to describe the numerical value of the ground state energy, we recall that for any $N\in\N$ there exists a uniquely determined triple\\
\mbox{$(s(N),k(N),j(N))\in(\N\cup\{0\})^3$} with $ k(N)\le 5$ and $ j(N)\le s(N)$ such that
\begin{equation*}
N=3s^2(N)+3s(N)+1+(s(N)+1)k(N)+j(N).
\end{equation*}
In terms of the numbers $s$, $k$, and $j$, it follows from \cite{HR, DLF} that 
\begin{equation}\label{minimformula}
\min_{X\in\XLN}P(X)=\left\{\begin{array}{ll}
6\,s(N)&\textrm{if }k(N)=j(N)=0,\\
6\,s(N)+k(N)+1&\textrm{otherwise}.
\end{array}\right.
\end{equation}
Finally, we will need the following result which is an immediate consequence of \cite[Lemma 4.1, Lemma 4.2]{DLF}.

\begin{lemma}\label{lemmaimp}
Let $N,N'\in\N$.

Let $X\in\XLN$ and let $X':=X\setminus \partial X$. Then
$$
P(X)\ge P(X')+6.
$$
Conversely, let $X'\in\XLNN$ and set
$$
X:=X'\cup\{x\in\mathcal L:\,\textrm{there exists }x'\in X'\textrm{ with }|x-x'|=1\};
$$
then 
$$
P(X)= P(X')+6.
$$
\end{lemma}

We are now in a position to prove the uniqueness statement in Theorem \ref{uniqueness}. 
We establish the claim only for $N$ of the form (b), the proof in case (a) being analogous (and an alternative proof in case (a) being already known \cite{S}).
We use induction on $s=s(N)$. It is easy to see that the claim is satisfied for $s=0,1$, for any value of $k$ in $\{0,1,2,3,4\}$. Fix such a $k$.
Assuming that the minimizer is unique for $N_{s-1}:=3(s-1)^2+3(s-1)+1+s\,k+s-1$, we prove that it is unique for $N_{s}:=3\,s^2+3\,s+1+(s+1)k+s$.
Let $X$ be a minimizer of $\E_{HR}$ among $N$-particle configurations; then by the Heitmann-Radin theorem $X$ belongs (up to rotation and translation) to $\XLN$, and by \eqref{minimformula} it satisfies $P(X)=6\,s+k+1$. Set $X':=X\setminus \partial X$. We have that $\sharp X'=N_s-6\,s-k-1=N_{s-1}$.
By Lemma \ref{lemmaimp}, we conclude that
$$
6\,s+k+1=P(X)\ge P(X')+6\ge \min_{Y\in\XLNN}P(Y)+6= 6\,(s-1)+k+1+6,
$$
and hence all the above inequalities are equalities. Therefore $X'$ is a minimizer. By the inductive assumption, the minimizer $X'$ is unique (up to rotation and translation). Since, by the construction in Lemma \ref{lemmaimp}, the set $X$ is fully determined by $X'$, we obtain the claim.

\section{Concluding remarks}

The above arguments show that
\begin{equation*}
 \{ N\ge 3\, : \, \mbox{the minimizer of }P\mbox{ in }\XLN \mbox{ is unique} \} = 
 \{ a_n \}_{n\ge 3},
\end{equation*}
where 
\begin{equation} \label{character0}
   a_{n}:=\left\{\begin{array}{ll}
   3\,s^2+3\,s+1 &\textrm{if }n=6\,s,\\
   3\,s^2+3\,s+1 +(s+1)k+s&\textrm{if }n=6\,s+k+1, \, k\in\{0,1,2,3,4\}. 
   \end{array}\right.
\end{equation}
Moreover, using \eqref{minimformula} one can easily check that for any $n\ge 3$, $a_n$ has a simple geometric meaning: it is the maximum particle number whose Heitmann-Radin minimizer has perimeter $n$, in formulae:

\begin{equation}\label{character}
a_n=\max\{N\in\N\,:\,\min_{X\in\XLN}P(X)=n\}.
\end{equation}

It appears that the sequence $\{a_n\}$ is well-known in number theory (see for instance \cite{PS, oeis}). For example, $a_n$ seems to co-incide with the number of nonnegative integer solutions to the Diophantine equation
$$
       x+2\,y+3\,z=n.
$$
Nevertheless, to the best of our knowledge, neither the geometric characterizations of $a_n$ in Theorem \ref{uniqueness} and \eqref{character} nor the explicit one given in \eqref{character0}  have been observed before. It would be very interesting both from a physical and a mathematical point of view if a direct argument linking the Heitmann-Radin model with the above Diophantine equation could be found. 
\\[3mm]

{\bf Acknowledgements:}
We are indebted to Michael Baake for pointing out to us the website \cite{oeis} after attending a lecture on the work reported here.  
The reesearch of LDL was funded, and that of GF partially supported, by the DFG Collaborative Research Center CRC 109 ``Discretization in Geometry and Dynamics''.

\end{document}